# The Bat Signal

Keith Zengel[1] and Anna Klales[2]
1. Wentworth Institute of Technology, Boston, MA. email: zengelk@wit.edu
2. Harvard University, Cambridge, MA.
(Dated: July 10, 2024)

Here is a simple experiment you can run remotely or in person with basic household items: making a bat signal. First, have your students cut out a bat shape from black construction paper, then have them hold the bat in front of a desk lamp aimed at the wall. This setup looks like what we see in the movies and comics, but it doesn't work! Instead of a projected bat shadow surrounded by a circle of light, we only see a slightly dimmer light projected onto the wall. With a little Socratic prodding, you can get your students to explore the basic principles of geometric optics to make functioning pinhole, pinpoint, and thin-lens bat signals.

The problem with the light-bat setup is that each point on the bulb acts as a light source that emits light rays in all directions. Portions of the bulb not blocked by the bat will still emit light that reaches the wall, as shown in Fig. 1. The result is a proportionally dimmer light on the wall, but no shadow.

Students may notice that they can project a bat shadow onto the wall if they hold the bat near the wall and far from the lamp. The only problem is that this is no way to project the bat signal onto a faraway surface (the night clouds over Gotham).

In order to project the bat shadow, they can use a light source smaller than the bat (smartphone flashlights work well) or place in front of the desk lamp another sheet of construction paper with a pinhole in it. If the items are arranged light-pinhole-bat, then a bat shadow is projected onto the wall. This works because the pinhole acts as a single point-source that emits light radially. The rays blocked by the bat produce a shadow and there is no other point-source to send light to the shadowed part of the wall.

Of course, students can also shuffle the order and try a light-bat-pinhole arrangement. This projects an inverted bat shadow onto the wall and gives you a chance to introduce the physics of pinhole cameras.[1, 2]

Somewhat remarkably, students can also use a *pinpoint* to project a bat shadow.[3] In this case, they place the "bat-hole" (the hole in the sheet from which they cut their bat) over the light and hold a small opaque point in front of the bat hole. We made our "pinpoints" by balling up a bit of aluminum foil on the end of a paperclip.

The pinpoint setup works with the shadow-logic of the pinhole: light is admitted everywhere *except* at a single point.

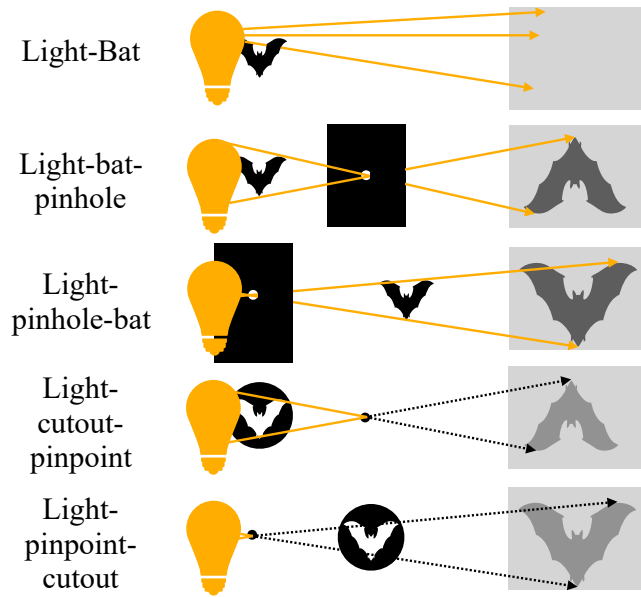

FIG. 1: Sketches of the different arrangements of light, pinhole, pinpoint, bat, and bat cutouts. In all cases the light is on the left and the wall is on the right. Note that in both pinpoint cases the shadow is fainter, as some light rays reach the bat shadow region on the wall unobstructed by the pinpoint.

So the pinpoint blocks the light rays that travel through that one point, resulting in an inverted bat shadow on the wall! A larger pinpoint will create a darker but blurrier shadow, while a smaller pinpoint will create a fainter but better defined shadow. Again, both the light-cutout-pinpoint and light-pinpoint-cutout arrangements result in bat shadows, with the former being inverted and the latter upright.

A fun experiment is to place the upside-down cutout near the light and the pinpoint near the wall. In this arrangement, students see the pinpoint shadow on the wall. Then, as they move the pinpoint closer to the cutout, the shadow grows wings and becomes a bat! They can see a similar effect by placing an upside-down bat near the light and the pinhole near the wall, then moving the pinhole closer to the bat.

Sketches of all the above arrangements are shown in Fig. 1.

Finally, students can use a magnifying glass to project a very crisp inverted bat signal far away by holding the magnifying glass a touch more than a focal length from the lamp. This places the image plane a touch less than infinity away, as the light rays emitted from the lens are nearly parallel. Many students will appreciate the striking demonstration of a magnifying glass projecting the text from the surface of a light bulb across the room.[4] You can then block those nearly parallel light rays with a bat cutout to create a distant bat shadow. Similarly, you could place a light source near the focal point of a concave mirror and block the outgoing parallel light rays with a bat cutout to project a shadow. The concave mirror setup is a homemade searchlight, similar to the bat signal setup students have seen in the movies.

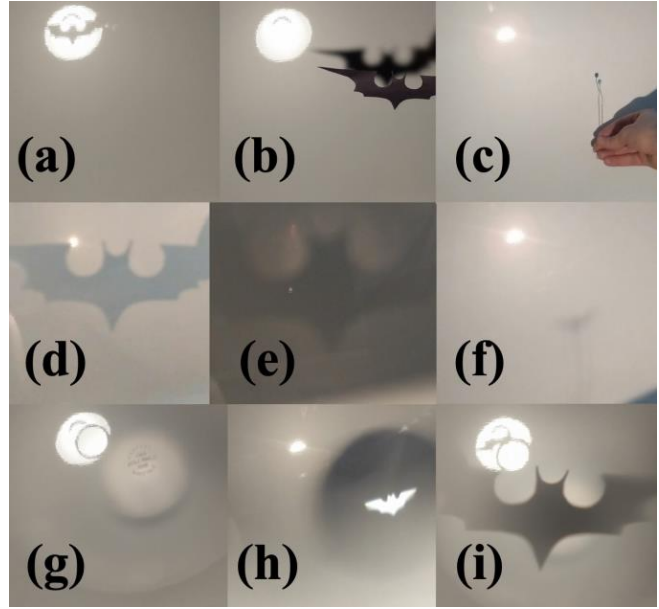

FIG. 2: Various bat signal setups projected onto a reflective white board. The reflection shows the light source and objects near it. In (a) we see the reflection of a bat in front of the bulb, but no projected shadow. In (b) and (c) we see a bat and pinhole near the board projecting shadows. In (d) we see the shadow from a light-pinhole-bat arrangement, while (e) shows the pinhole image of an inverted bat in a light-bat-pinhole arrangement. In (f) we see the shadow of an inverted bat cutout in a light-cutout-pinpoint arrangement. The light-pinpoint-cutout shadow was too faint for a photograph, but is visible in person. In (g), (h), and (i) we see the images produced by a bare light bulb, an inverted bat, and an inverted bat cutout produced by a magnifying glass held a focal length from the bulb.

Photos of our experiments are shown in Fig. 2. If your students don't like bats or batman, they can use these techniques to project any shape they like!

We used this lab in an algebra-based physics course for non-STEM majors as part of an early lab unit on vectors, though it also serves as a nifty introduction to geometric optics and potentially as a fun project for K-12 outreach.